\documentclass[conference]{IEEEtran}
\IEEEoverridecommandlockouts

\usepackage[a4paper]{geometry}
\newgeometry{left=1.35cm,right=1.35cm,bottom=1.35cm,top=1.35cm}

\usepackage{setspace}
\usepackage{enumitem}
\usepackage{amsmath} 
\usepackage{cite}
\usepackage{amsmath,amssymb,amsfonts}
\usepackage[ruled,vlined]{algorithm2e}
\usepackage{algorithmic}
\usepackage{graphicx}
\usepackage{textcomp}
\usepackage{caption}
\usepackage{subcaption}
\usepackage{xcolor}

\def\BibTeX{{\rm B\kern-.05em{\sc i\kern-.025em b}\kern-.08em
    T\kern-.1667em\lower.7ex\hbox{E}\kern-.125emX}}

\usepackage{fancyhdr}
\renewcommand{\IEEEbibitemsep}{0.5pt}
\makeatletter

\IEEEtriggercmd{\reset@font\normalfont\fontsize{6.8pt}{7.11pt}\selectfont}
\makeatother
\IEEEtriggeratref{1}
    
\begin{document}

\title{CoNLoCNN: Exploiting Correlation and Non-Uniform Quantization for Energy-Efficient Low-precision Deep Convolutional Neural Networks\vspace{-10pt}}

\author{\IEEEauthorblockN{Muhammad Abdullah Hanif\textsuperscript{ 1,2}, Giuseppe Maria Sarda\textsuperscript{ 3}, Alberto Marchisio\textsuperscript{ 2},\\Guido Masera\textsuperscript{ 3}, Maurizio Martina\textsuperscript{ 3}, Muhammad Shafique\textsuperscript{ 1}}
\IEEEauthorblockA{
\textsuperscript{1}\textit{New York University Abu Dhabi,} Abu Dhabi, UAE \\
\textsuperscript{2}\textit{Technische Universität Wien (TU Wien)}, Vienna, Austria \\
\textsuperscript{3}\textit{Politecnico di Torino}, Turin, Italy \\
mh6117@nyu.edu, s255648@studenti.polito.it, alberto.marchisio@tuwien.ac.at,\\guido.masera@polito.it, maurizio.martina@polito.it, muhammad.shafique@nyu.edu}
\vspace{-25pt}}

\maketitle
\thispagestyle{fancy}
\begin{abstract}
In today's era of smart cyber-physical systems, Deep Neural Networks (DNNs) have become ubiquitous due to their state-of-the-art performance in complex real-world applications. The high computational complexity of these networks, which translates to increased energy consumption, is the foremost obstacle towards deploying large DNNs in resource-constrained systems. Fixed-Point (FP) implementations achieved through post-training quantization are commonly used to curtail the energy consumption of these networks. However, the uniform quantization intervals in FP restrict the bit-width of data structures to large values due to the need to represent most of the numbers with sufficient resolution and avoid high quantization errors. In this paper, we leverage the key insight that (in most of the scenarios) DNN weights and activations are mostly concentrated near zero and only a few of them have large magnitudes. We propose CoNLoCNN, a framework to enable energy-efficient low-precision deep convolutional neural network inference by exploiting: (1) non-uniform quantization of weights enabling simplification of complex multiplication operations; and (2) correlation between activation values enabling partial compensation of quantization errors at low cost without any run-time overheads. To significantly benefit from non-uniform quantization, we also propose a novel data representation format, Encoded Low-Precision Binary Signed Digit, to compress the bit-width of weights while ensuring direct use of the encoded weight for processing using a novel multiply-and-accumulate (MAC) unit design.
\end{abstract}

\section{Introduction}

Deep Neural Networks (DNNs) are state-of-the-art models for applications like object classification, image segmentation and speech processing~\cite{lecun2015deep}. However, they have high computational complexity and memory footprint, which translates to high hardware and energy requirements~\cite{sze2017efficient}. \textit{This resource-hungry nature of DNNs challenges their high-accuracy deployment in resource-constrained scenarios such as inference on resource-constrained embedded devices.} 

Methods like pruning and quantization are used to reduce the computational complexity and memory footprint of DNNs~\cite{anwar2017structured, han2015deep, zhu2016trained, rastegari2016xnor}. However, \textit{these state-of-the-art approaches are highly effective when used with retraining} to minimize the accuracy loss. While retraining smaller DNNs designed for less complex datasets incurs fewer overheads, it is super costly to retrain larger DNNs designed for complex applications, which can take several days even on high-end GPU servers. 
Moreover, proper retraining of DNNs during optimization phase requires access to a comprehensive dataset, and may not be possible in cases where the dataset is an IP of a company and it has not made it available (e.g., Google’s JFT-300M dataset~\cite{sun2017revisiting}) to the end user who wants to optimize the given pre-trained DNN for a specific set of resource constraints.
Under such cases, an effective approach is post-training quantization that enables low-precision Fixed-Point (FP) implementation ($\leq$8 bits) of DNNs~\cite{gysel2016hardware}\cite{lin2016fixed}, and thereby reduces power, latency, and memory footprint. 
However, reducing precision introduces quantization errors, leading to potentially noticeable accuracy loss.  
\textit{This additional source of error degrades the application-level accuracy of DNNs and restricts the designers from reducing the bit-widths of DNN datastructures beyond a level without significantly affecting the accuracy.}

\textbf{State-of-the-Art Works and Their Limitations:} To overcome the above-mentioned limitations and achieve significant efficiency gains, several works have been proposed. 
Compensated-DNN~\cite{jain2018compensated} proposed a technique to quantize DNN data-structures and dynamically compensate for quantization errors. It achieves this using a novel Fixed Point with Error Compensation (FPEC) data representation format and a specialized Processing Element (PE) design containing a low-cost error compensation unit. 
In FPEC, each number consists of two sets of bits: (1) bits that represent quantized value of the number in traditional FP formats; and (2) bits that store quantization related information, such as error direction, which is useful for error compensation. 
Techniques like~\cite{tann2017hardware} and~\cite{jain2019biscaled} make use of power-of-two quantization and multi-scaled FP representation (respectively) to reduce the bit-width of DNN data-structures as well as the complexity of the PEs (mainly multipliers). Recently, CAxCNN~\cite{9137167} proposed the use of reduced precision Canonical-Sign-Digit (CSD) representation to decrease the complexity of Multiply-and-Accumulate (MAC) units in DNN accelerators. 
Similarly, \cite{trinh2015efficient} combined the use of reduced precision CSD with customized delta encoding scheme to efficiently approximate and encode DNN parameters. A summary of the key characteristics of the most relevant state-of-the-art techniques is presented in Table~\ref{tab:comparison_table}. 
\textit{The table shows that Compensated-DNN is the only work that employs error compensation (without retraining) to offer better energy-accuracy trade-off. 
However, it does not exploit the advantages of efficient data representations such as power-of-two or reduced precision CSD representations. Moreover, it also requires extra hardware support for error compensation.} 

\begin{table}[h] 
\centering
\caption{Characteristics of the most relevant state-of-the-art works. RP-CSD refers to reduced precision CSD and T-CSD refers to truncated CSD representation. }  
\includegraphics[width=1\linewidth]{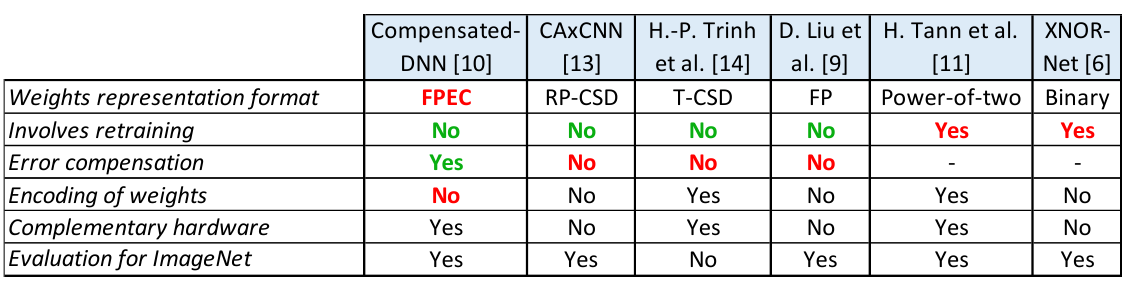}\vspace{-20pt}
\label{tab:comparison_table}
\end{table}

Other approaches that are also designed to improve the energy efficiency of DNN inference without the need of retraining involve the use of approximate hardware modules (mainly approximate multipliers)~\cite{7298218}\cite{mrazek2019alwann}, and voltage-scaling of computational array and memory modules~\cite{zhang2018thundervolt}\cite{kim2018matic}. 
AxNN~\cite{7298218} selectively approximates the neurons that do not impact the application-level accuracy much. However, it offers close to the baseline accuracy only with approximation-aware retraining. 
ALWANN~\cite{mrazek2019alwann} presents a method that employs functional approximations in the MAC units to improve the energy efficiency of DNN inference without involving approximation-aware retraining. 
It mainly determines the most suitable approximation for each layer of the given DNN to achieve a reasonable amount of savings. 
\textit{However, note that these techniques have shown effectiveness only at 8-bit FP precision level (i.e., not for less than 8-bit precision) and only for relatively less complex datasets like Cifar-10.} 
Moreover, the voltage scaling techniques such as ThunderVolt~\cite{zhang2018thundervolt} and MATIC~\cite{kim2018matic} either introduce faults at run-time that can lead to undesirable accuracy loss or the need for expensive fault-aware retraining. 

\textbf{Summary of Key Challenges Targeted in this Work:} Based on the above-mentioned limitations of the existing works, we highlight the following key challenges: 
\begin{itemize}[leftmargin=*]
    \item There is a need to investigate existing DNN post-training quantization and approximation techniques to identify the methods and data representation formats that can enable high energy and performance efficiency without affecting the application-level accuracy of DNNs. 
    \item To maintain high accuracy, dynamic error compensation requires additional resources that increase DNN inference cost. Therefore, there is a need to explore low-cost software-level error compensation mechanisms that are required to be applied only once during DNN conversion (i.e., design-time phase) and have the potential to offer benefits equivalent to costly dynamic error compensation techniques.  
\end{itemize}

\begin{figure}[htbp]
\centering
\includegraphics[width=1\linewidth]{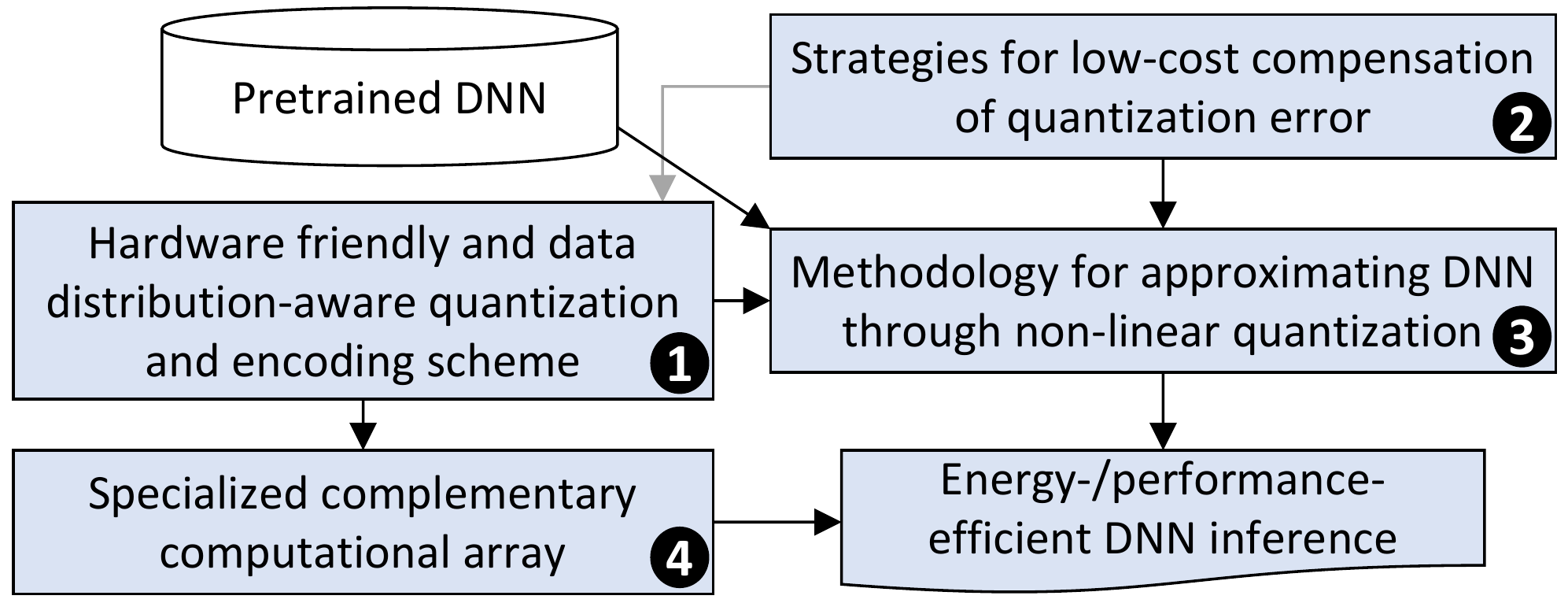}
\caption{System overview with our novel contributions and flow}\vspace{-10pt}
\label{fig:novel_contributions}
\end{figure}

\textbf{Our Novel Contributions:} We propose an algorithm and architecture co-design framework, CoNLoCNN, for effectively approximating DNNs through post-training quantization to improve the power-/energy-efficiency of DNN inference process without involving retraining. Towards this:

\begin{enumerate}[leftmargin=*]
    \item We investigate different methods for quantizing DNN data-structures in search for an effective method for approximating DNNs that offers high energy-accuracy trade-off. We identify that choosing a data representation format that is aligned to the long-tailed data distribution of DNN parameters (see Fig.~\ref{fig:data_distribution}) results in less overall quantization error, and this selection can also help in simplifying the hardware components (mainly multipliers in the processing array of DNN hardware accelerators) (\textbf{Section~\ref{sec:data_distribution}}). Based on our analysis, we propose an \textit{Encoded Low-Precision Binary Signed Digit} data representation format that at the core is an evolved version of power-of-two representation and thereby enables the use of low-cost shift operations instead of multiplications. This results in significant simplification of the MAC units in DNN accelerators and thereby enables high energy and performance efficiency. (\textbf{Section~\ref{sec:novel_number_representation}})
    \item We propose a low-cost error compensation strategy that is required to be applied only once at the conversion-time (compile-time) to compensate for quantization errors. This enables the users to completely avoid overheads associated with dynamic compensation. (\textbf{Section~\ref{sec:novel_correlation_strategy}})
    \item We propose a systematic methodology for quantizing DNNs while exploiting the proposed error compensation strategy. (\textbf{Section~\ref{sec:novel_methodology_overall}})
    \item We propose a specialized multiply-and-accumulate (MAC) unit design that fully exploits the potential of the proposed approximation scheme for improving the energy efficiency of DNN inference. (\textbf{Section~\ref{sec:novel_hardware_design}})
\end{enumerate}
A summary of our novel contributions is shown in Fig.~\ref{fig:novel_contributions}.

\section{Preliminaries}

This section introduces the key terms used in this paper.

\subsection{Deep Neural Networks (DNNs)}
A DNN is an interconnected network of neurons, where each neuron performs a weighted sum operation, and then passes the output through a non-linear activation function. The functionality of a neuron can mathematically be written as $O = F(\sum_i W_i * A_i + b)$, where $O$ represents the output, $W_i$ represents the $i^{th}$ weight, $A_i$ represents the $i^{th}$ input (a.k.a. activation), $b$ represents the bias, and $F$ represents the non-linear activation function. In DNNs, neurons are arranged in the form of layers. The layers are then connected in a specialized format to form a DNN. 

\subsection{Convolutional Neural Networks (CNNs)}

\begin{figure}[ht]
\centering
\includegraphics[width=1\linewidth]{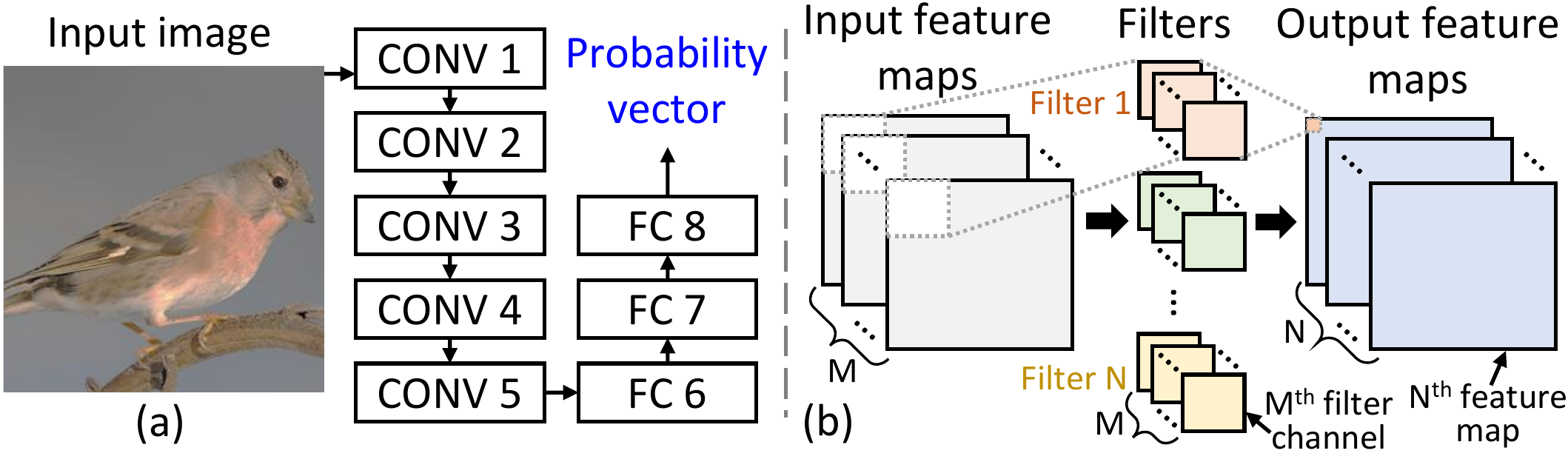}
\caption{(a) Architecture of AlexNet. (b) Illustration of a CONV.\vspace{-10pt}}
\label{fig:2}
\end{figure}

A CNN is a type of DNN specialized for processing spatially correlated data such as images~\cite{lecun2015deep}. It is mainly composed of convolutional (CONV) layers and fully-connected (FC) layers (see Fig.~\ref{fig:2}(a)). The CONV layers are used to extract features from the input by convolving it with a set of filters (see Fig.~\ref{fig:2}(b)). 
The output generated by convolving a filter with the input is referred as a feature map. The FC layers are typically used at the end of the network for classification. A FC layer is composed of several neurons where each neuron receives the complete output vector of the previous layer to generate an output. 

\section{Strategies for Enabling Low-Precision and Energy-Efficient DNN Inference}
\label{sec:analysis_and_strategies}

\subsection{Strategy 1: Select a quantization scheme that is aligned with the data distribution} \label{sec:data_distribution}

\begin{figure}[h]
     \centering
     \begin{subfigure}[h]{0.49\linewidth}
         \centering
         \includegraphics[width=\linewidth]{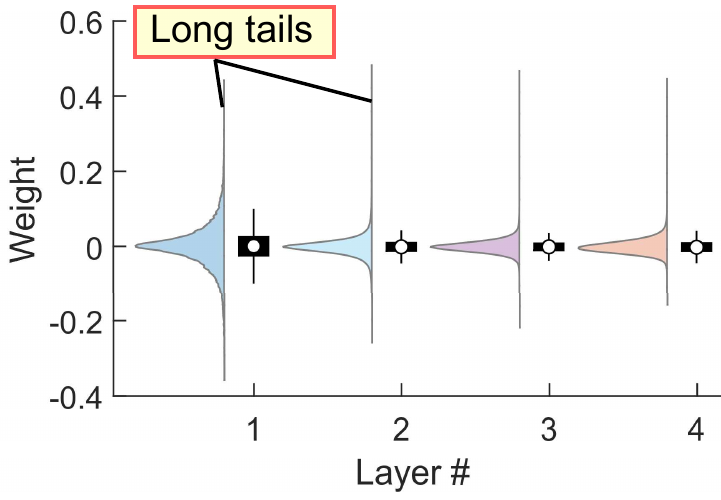}
         \caption{Distribution of weights of the layers}
     \end{subfigure}
     \hfill
     \begin{subfigure}[h]{0.49\linewidth}
         \centering
         \includegraphics[width=\linewidth]{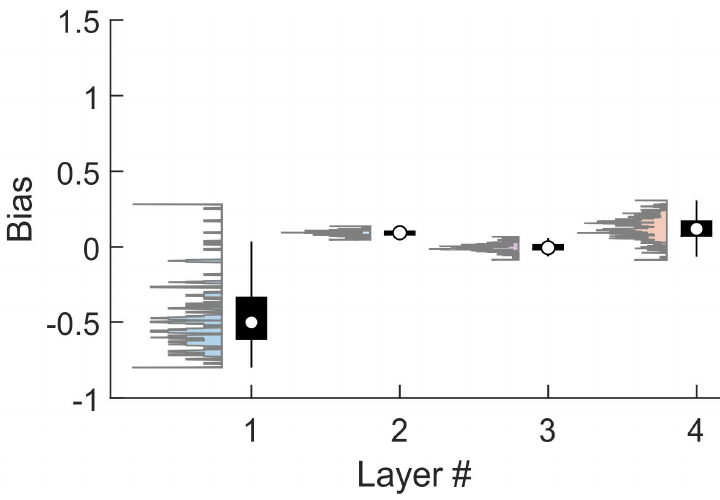}
         \caption{Distribution of biases of the layers}
     \end{subfigure}
     \vfill
     \begin{subfigure}[h]{1\linewidth}
         \centering
         \includegraphics[width=\linewidth]{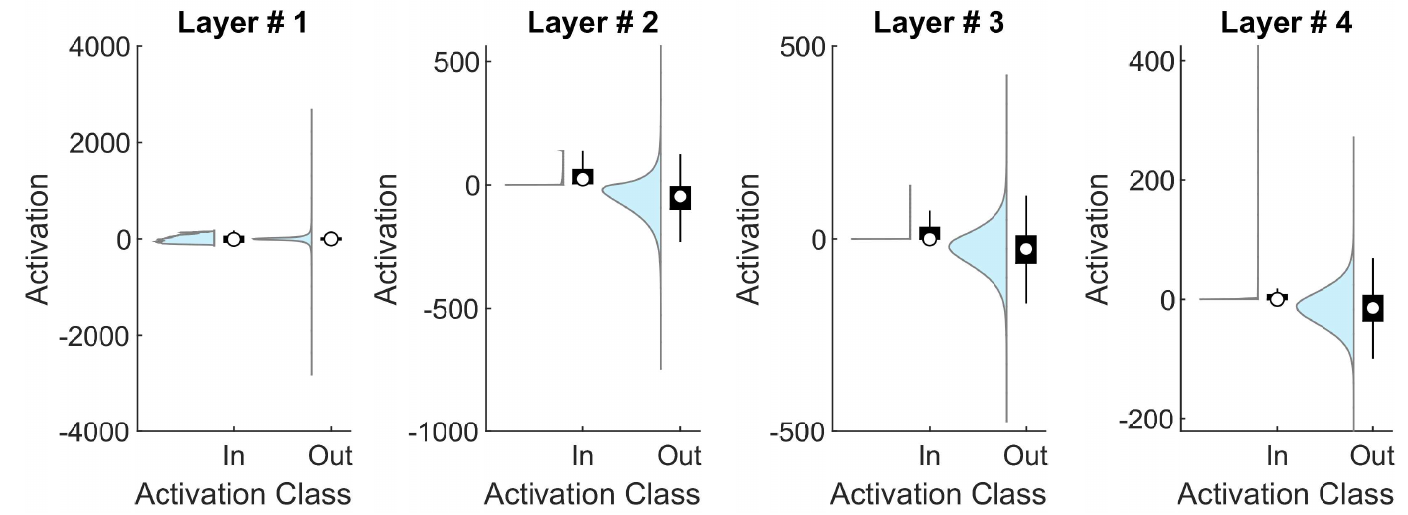}
         \caption{Distribution of input and output activations of the layers}
     \end{subfigure}
        \caption{Distribution of weights, biases and activations of the first four convolutional and layers of the AlexNet (in the form of half-violin plots and box plots). Note that the output activations here represent the output of the layer before passing through activation functions.\vspace{-10pt}}
        \label{fig:data_distribution}
\end{figure}

\begin{figure}[htbp]
\centering
\includegraphics[width=1\linewidth]{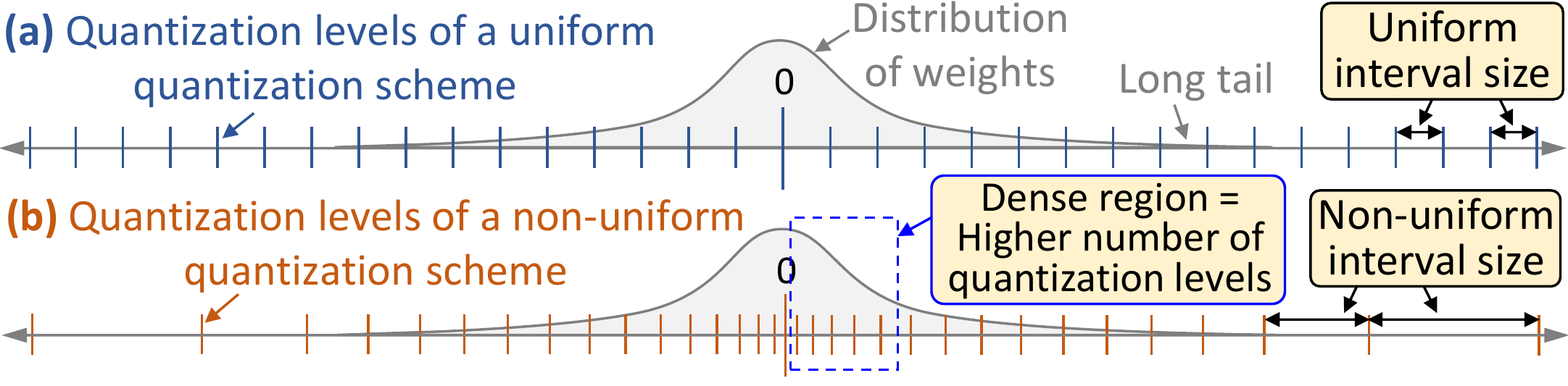}
\vspace{-10pt}
\caption{Comparison between uniform and non-uniform quantization.\vspace{-10pt}}
\label{fig:Quant_comp}
\end{figure}

Fig.~\ref{fig:data_distribution} shows the distributions of weights, biases, and activations of the layers of a trained AlexNet. 
It can be observed from the figure that the distributions of weights and activations have long tails, i.e., in each data-structure, majority of the values have small magnitude and only a limited number of values have large magnitude. Moreover, for each layer, the distribution of weights is close to a Gaussian distribution with mean equal to zero. Considering the data distributions, a low precision uniform quantization (for example, see Fig.~\ref{fig:Quant_comp}(a)) 
would result in high overall quantization error compared to a non-uniform quantization (for example, see Fig.~\ref{fig:Quant_comp}(b)) 
having the same (or less) number of quantization levels, assuming the levels are distributed based on the data distribution, i.e., more number of narrowly-spaced quantization levels in dense regions and less number of widely-spaced quantization levels in light regions. 
Therefore, aligning quantization scheme with the data distribution can help in reducing the overall/average quantization error. However, a potential limitation of this is that, in case of low-precision non-uniform quantization, it can result in a significant increase in the maximum quantization error leading to high error variance. \textit{Hence, the ideal quantization scheme should balance between the average and the maximum quantization error to achieve minimum quality degradation.} 

\subsection{Strategy 2: Exploit correlation between neighboring feature map values to reduce the effective variance and mean of quantization error}
\label{sec:novel_correlation_strategy}
Here, we first analyze the impact of variations in the bias values of a CNN on its classification accuracy. 
Note that this analysis mainly helps us understand the effects of errors that affect the mean of activation values. 
Then, we study the correlation of data within and across input feature maps of a layer to see if quantization errors can be partially compensated by distributing them across weights in the same filter/neuron. 
Afterward, we present a mathematical analysis and show how the gained insights can be exploited to reduce the impact of quantization errors. 

\subsubsection{Impact of modifying the bias values in DNNs}

\begin{figure}[htbp]
\centering
\includegraphics[width=1\linewidth]{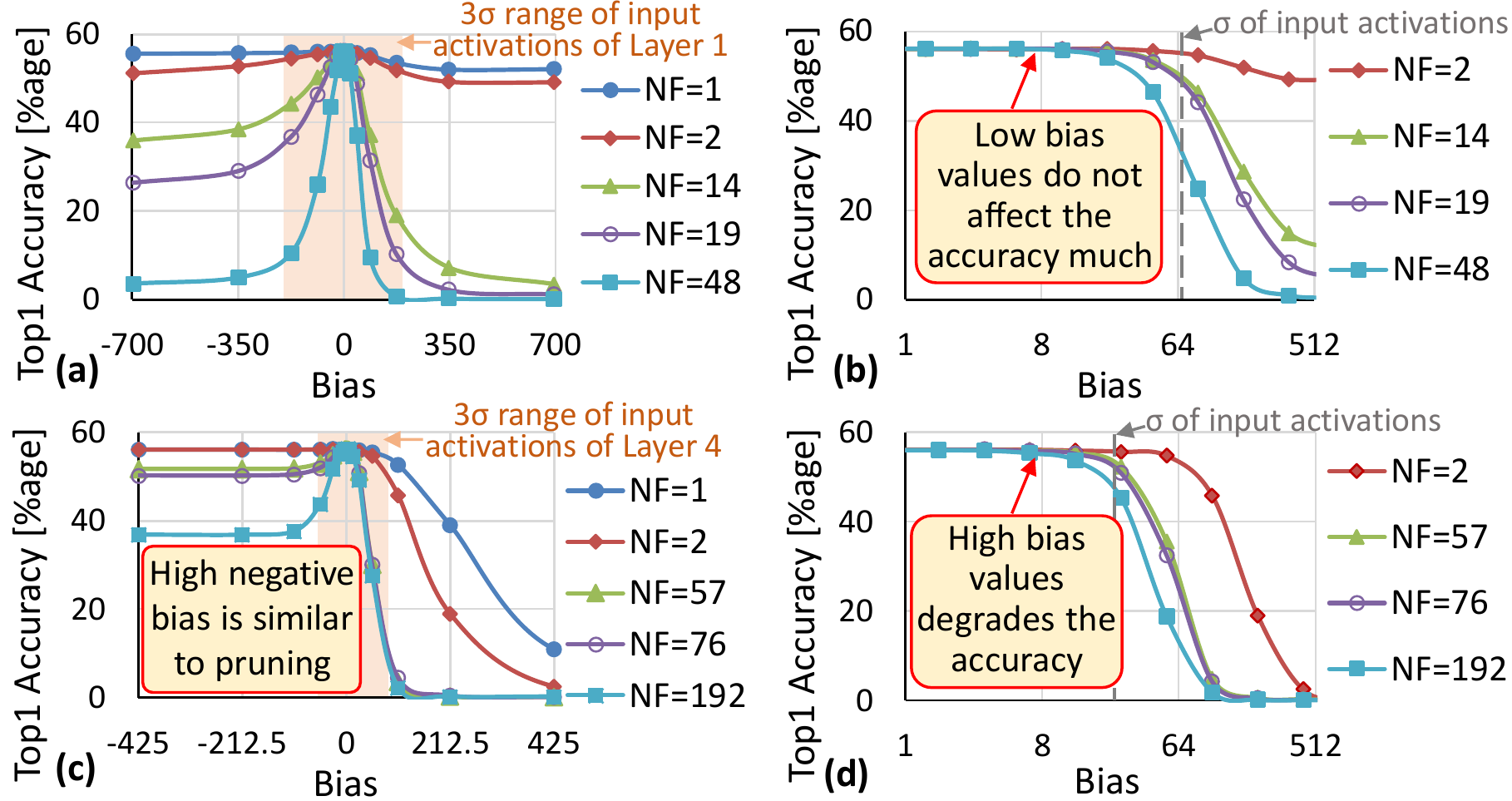}
\caption{Impact of altering the bias values of different number of randomly selected filters/neurons (NF) of different layers of a trained AlexNet (trained on the ImageNet dataset) on its classification accuracy. (a) and (c) show the impact when same amount of positive (or negative) value is added to the bias values of the selected filters of layer 1 and 4, respectively. (b) shows the impact when the bias values of half of the selected filters of layer 1 are injected with positive noise and half with negative noise having the same magnitude. Similar to (b), (d) shows the results for layer 4.\vspace{-5pt}}
\label{fig:Bias_AlexNet_Mean}
\end{figure}

Fig.~\ref{fig:Bias_AlexNet_Mean} shows the impact of varying the bias of different number of filters of a layer of a pre-trained AlexNet on its classification accuracy. From Fig.~\ref{fig:Bias_AlexNet_Mean}(a) and Fig.~\ref{fig:Bias_AlexNet_Mean}(c) it can be observed that, when a small constant value, i.e., a value close to the range of original bias values (see Fig.~\ref{fig:data_distribution}(b)), is added to the bias values of a number of filters, the accuracy of the network stays close to its baseline. However, when the magnitude of the constant is large, it degrades the accuracy. 
The difference between the impact of positive and negative noise is mainly due to the presence of ReLU activation function in the AlexNet, as a large negative bias leads to a large negative output which is then mapped to zero by the ReLU function and thereby limits the impact of error on the final output. 
Fig.~\ref{fig:Bias_AlexNet_Mean}(b) and Fig.~\ref{fig:Bias_AlexNet_Mean}(d) show that, when the bias values of half of the filters are injected with positive noise and half with negative noise (all having the same magnitude), the behavior of classification accuracy is dominated by the behavior of filters that are injected with positive noise. 

\begin{figure}[htbp]
\centering
\includegraphics[width=1\linewidth]{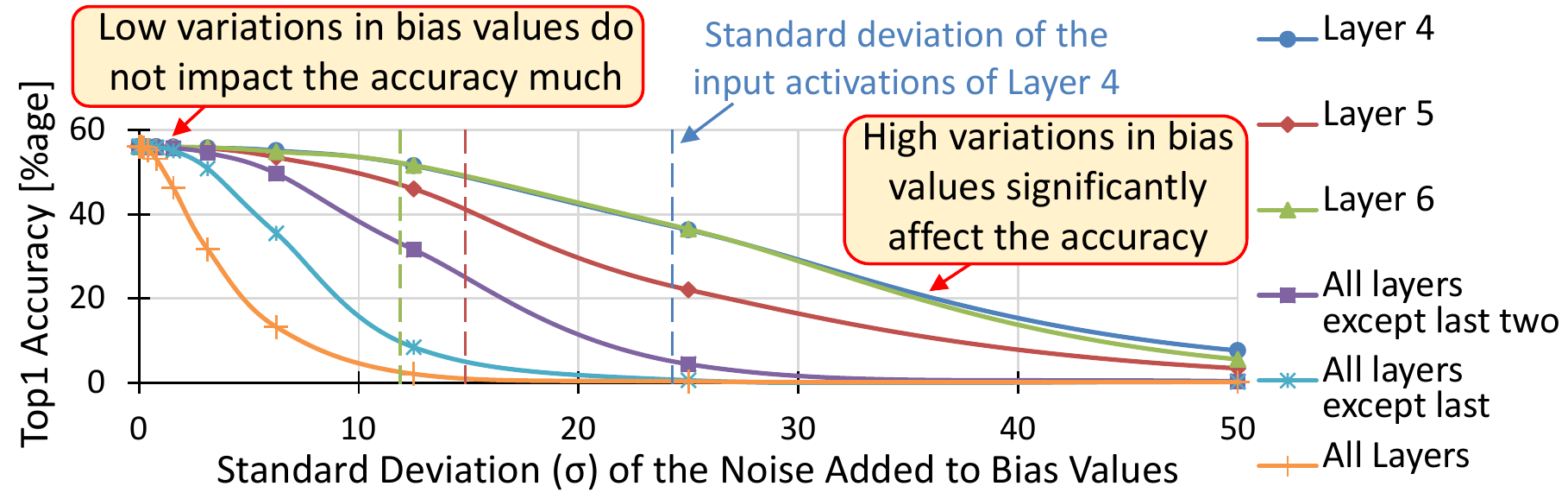}
\caption{Impact of adding noise generated using a Gaussian distribution to the bias values of filters/neurons of different layers and different number of layers of a trained AlexNet on its classification accuracy.\vspace{-5pt}}
\label{fig:Bias_AlexNet_SD}
\end{figure}

To further study the impact of mean shifts in the output feature maps, we performed an experiment where we added noise generated using a Gaussian distribution to the bias values of the filters/neurons of different layers of a pre-trained AlexNet (see Fig.~\ref{fig:Bias_AlexNet_SD}). We observed that when the noise is generated using smaller standard deviation values and is injected to the intermediate layers of the network, it does not impact the accuracy much. However, if the noise is injected to the last layer of the network or is generated using larger standard deviation values, it leads to a significant drop in the DNN accuracy. 

From the above analysis, we deduce the following three conclusions. \textit{(1) Small to moderate level of variations due to any error/noise source in the mean of activation values of all the layers except the last layer do not impact the accuracy much. (2) Mean shift in the output degrades the accuracy only if it is large in magnitude or it is in the output of last layer of a DNN. (3) Resilience of DNNs to small-to-moderate level of errors in bias values points to the significance of large activation values.}

\subsubsection{Correlation between activation values of input feature maps} 

\begin{figure}[htbp]
\centering
\includegraphics[width=1\linewidth]{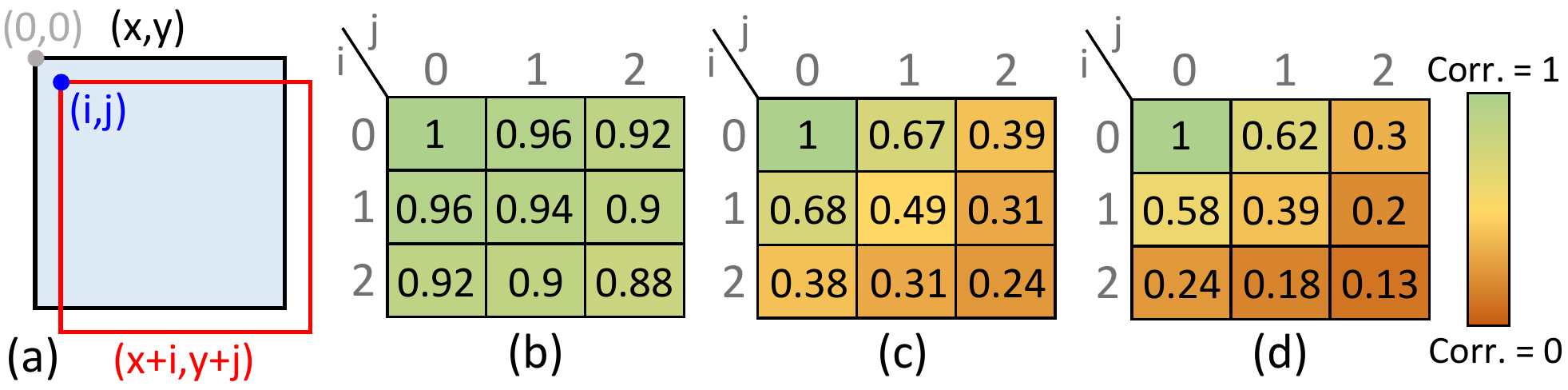}
\caption{Intra-feature map correlation of input activations of different layers of the AlexNet and the VGG16. (a) Illustration of an input feature map (shown in blue) and its shifted variant (shown with red border). $i$ and $j$ define the shifts in $x$ and $y$ directions, respectively. (b) and (c) show the correlation between the input feature maps and their shifted variants of layer 1 and layer 3 of the AlexNet, respectively. (d) shows correlation between neighboring input activations of layer 12 of the VGG16.\vspace{-5pt}} 
\label{fig:Intra_channel_corr}
\end{figure}

\textit{Intra-Feature Map Correlation:} Fig.~\ref{fig:Intra_channel_corr} shows the correlation between neighboring input activation values located at a constant shift from each other (represented using $i$ and $j$ in the figure) within input feature maps of a convolutional layer. Based on the correlation values in the figure, it can be said that, in all the convolutional layers of a DNN, there is a significant correlation between the neighboring activation values.  

\begin{figure}[h]
     \centering
     \begin{subfigure}[h]{0.33\linewidth}
         \centering
         \includegraphics[width=\linewidth]{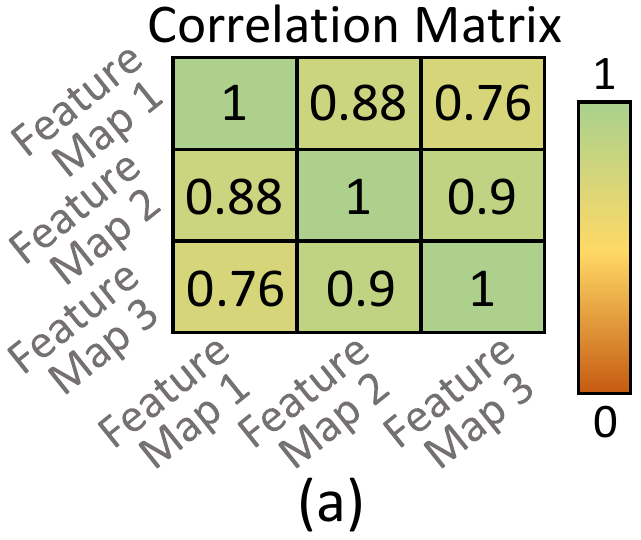}
     \end{subfigure}
     \hfill
     \begin{subfigure}[h]{0.65\linewidth}
         \centering
         \includegraphics[width=\linewidth]{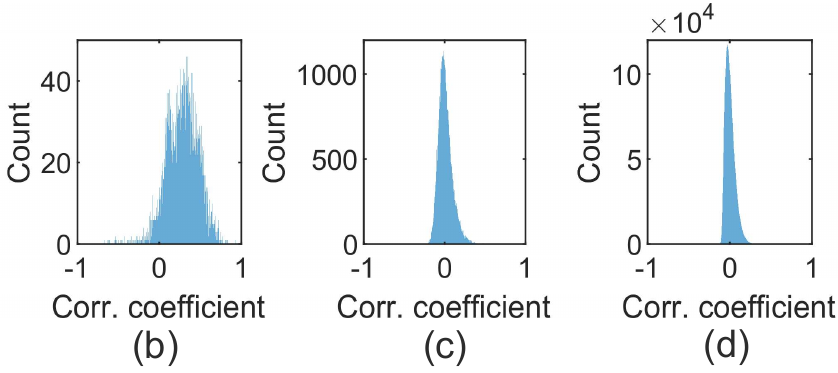}
     \end{subfigure}
        \caption{Correlation between input feature maps of different layers of the AlexNet. (a) Correlation matrix of input feature maps of layer 1. (b) Distribution of the correlation between input feature maps of layer 2. Similar to (b), (c) and (d) show distribtions of layer 4 and layer 7, respectively.\vspace{-10pt}}
        \label{fig:Inter_channel_corr}
\end{figure}

\textbf{Inter-Feature Map Correlation:} Fig.~\ref{fig:Inter_channel_corr} shows the distribution of correlation between different input feature maps of a layer of a pre-trained AlexNet. 
Fig.~\ref{fig:Inter_channel_corr}(a) shows that there is a significant correlation between the input feature maps of layer 1 of the network. The distributions in Figs.~\ref{fig:Inter_channel_corr}(b),~\ref{fig:Inter_channel_corr}(c), and~\ref{fig:Inter_channel_corr}(d) show that as we move deeper into the network the across-feature map correlation moves towards zero. 

\textit{Based on the above analysis, we conclude that only intra-feature map correlation can be exploited for error compensation.}

\subsubsection{Analysis of quantization error}

To analyze the effects of quantization on the output quality, let us consider a scenario in which we quantize the weights of a layer of a DNN and keep the activations in full-precision format. A quantized weight can be written as

\begin{equation}
    W_q = W + \Delta W 
\end{equation}

where $W$ represents unquantized weight and $\Delta W$ represents quantization error. If we assume $W ~ \sim \mathcal{N}(\mu_W,\,\sigma^{2}_W)$ and $\Delta W ~ \sim \mathcal{N}(\mu_{\Delta W},\,\sigma^{2}_{\Delta W})$, and $W$ and $\Delta W$ to be independent, then 

\begin{equation}
    W_q ~ \sim \mathcal{N}(\mu_{W} + \mu_{\Delta W},\,\sigma^{2}_{W} + \sigma^{2}_{\Delta W})
\end{equation} 

Similar to the distribution of $W$, for activations, we assume   

\begin{equation}
    A ~ \sim \mathcal{N}(\mu_A,\,\sigma^{2}_A)
\end{equation}

Now, for $O_q = \sum_{i = 1}^{n} W_{qi} * A_{i}$, using the above equations and assuming the weights and activations to be independent, we get 

\begin{multline}
    O_q ~ \sim \mathcal{N}(n * \mu_{A} * (\mu_{W} + \mu_{\Delta W}) ,\, n * ((\sigma^{2}_{W} + \sigma^{2}_{\Delta W} + \\ (\mu_{W} + \mu_{\Delta W})^2) * (\sigma^{2}_{A} + \mu_{A}^2) - \mu_{A}^2 * (\mu_{W} + \mu_{\Delta W})^2))
\end{multline}

As highlighted in the earlier analysis, small deviations in the mean of output activations do not impact the accuracy much; therefore, we mainly focus on comparing the variance of $O_q$ with the variance of $O = \sum_{i = 1}^{n} W_{i} * A_{i}$. Subtracting variance of $O$ from the variance of $O_q$, we are left with

\begin{multline}
    n * (\sigma^{2}_{\Delta W}*\sigma^{2}_{A} + \sigma^{2}_{\Delta W} * \mu_{A}^2 + \sigma^{2}_{A} * \mu_{\Delta W}^2 + 2 * \sigma^{2}_{A} * \mu_{W} * \mu_{\Delta W})
\end{multline}

\begin{figure}[htbp]
\centering
\includegraphics[width=1\linewidth]{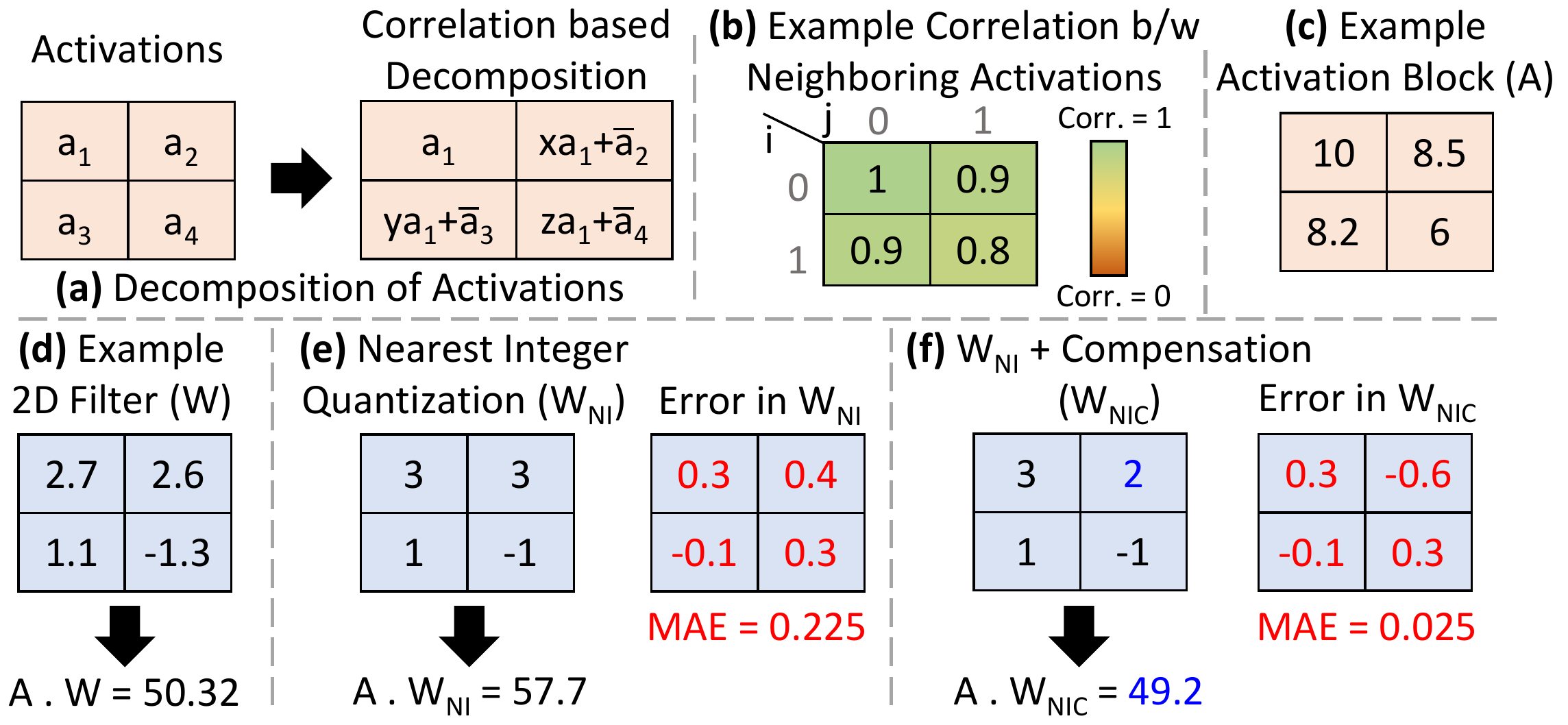}
\caption{Decomposition of activations, and an example illustrating the impact of exploiting correlation for error compensation on the output of dot-product operation.\vspace{-5pt}} 
\label{fig:corr_example}
\end{figure}

Now, to reduce the intensity of this additional term, we need to reduce the intensity of $\sigma_{\Delta W}$ and $\mu_{\Delta W}$. We can achieve this by exploiting the high correlation among the neighboring activation values in feature maps. 
Fig.~\ref{fig:corr_example}(a) shows a possible way of decomposing activation values of a block of feature map based on correlation among neighboring values. In case of high correlation (for example, see Fig.~\ref{fig:corr_example}(b)), the variables $x$, $y$ and $z$ (in Fig.~\ref{fig:corr_example}(a)) have high values (i.e., close to 1). 
As $\Bar{a_2}$, $\Bar{a_3}$ and $\Bar{a_4}$ represent the elements of $a_2$, $a_3$ and $a_4$ (respectively) that are orthogonal to $a_1$, the variables $\Bar{a_2}$, $\Bar{a_3}$ and $\Bar{a_4}$ exhibit low variance compared to $a_2$, $a_3$ and $a_4$ in the case of high correlation with $a_1$. 
We can exploit the presence of $xa_1$, $ya_1$ and $za_1$ in the neighboring activations to partially compensate/balance the error introduced in the dot-product of weights and activations due to quantization of weights by modifying the quantization scheme in such a way that it balances the mean quantization error of the neighboring weights. 

\textbf{Example:} To understand this, consider the activation block A shown in Fig.~\ref{fig:corr_example}(c) and 2D filter W shown in Fig.~\ref{fig:corr_example}(d). The output of dot-product of A and W comes out to be 50.32. Now, if we quantize the weights of the filter to the nearest integer values and perform the dot-product operation, we get 57.7 as the output; see Fig.~\ref{fig:corr_example}(e). 
However, if we map the value of the second weight to its other nearest integer value, i.e., 2 instead of 3 (see Figs.~\ref{fig:corr_example}(e) and~\ref{fig:corr_example}(f)), we can reduce the mean absolute error (MAE) in weights from 0.225 to 0.025 and the absolute error in the output of dot-product from 7.38 to 1.12. 
This shows that high correlation among the neighboring values enable us to reduce effective mean and variance of $\Delta W$ by minimizing the local mean quantization error inside filter channels. 

\subsubsection{Impact of Adjusting Intra-channel Mean Quantization Error in Weights}

\begin{figure}[htbp]
\centering
\includegraphics[width=1\linewidth]{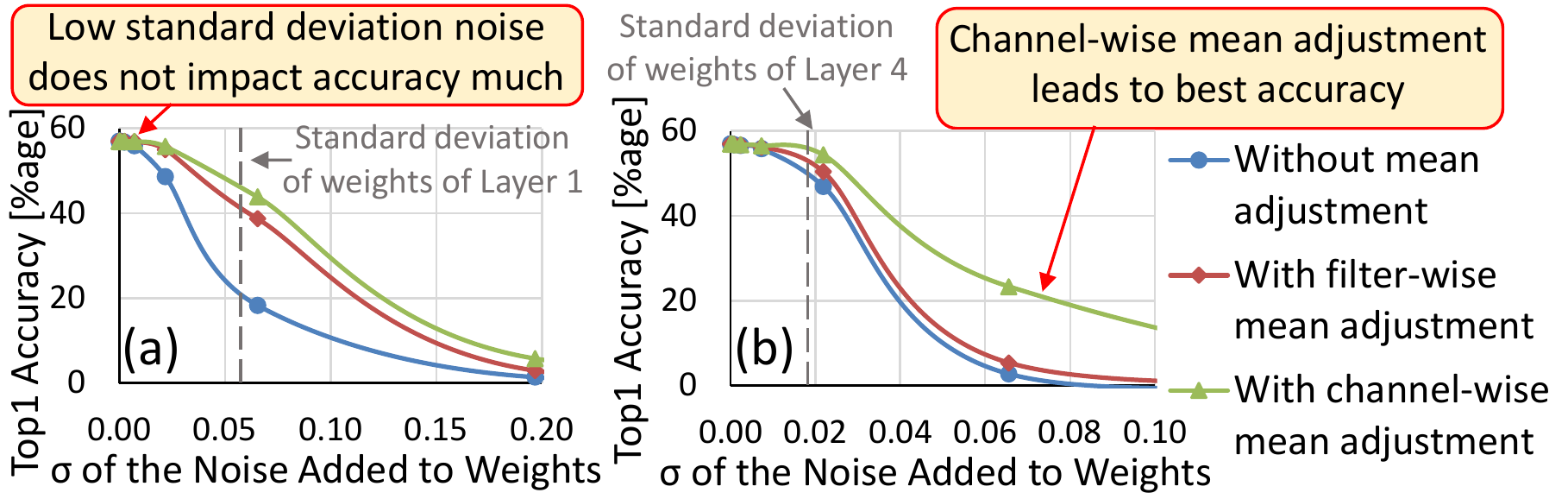}
\caption{Impact of adjusting the inter- and intra-channel mean quantization error in the weights of AlexNet. (a) Layer 1; (b) Layer 4.\vspace{-5pt}}
\label{fig:Weights_Noise_Mean_Error_Impact}
\end{figure}

To study the impact of adjusting the mean error in filters, we performed an experiment where we injected noise generated using a Gaussian distribution to the weights of the filters of layer~1 and layer~4 of the AlexNet. We studied three different cases: (1) No adjustment in the mean error of the weights; (2) Mean error adjustment case~1, where the mean error of each filter is subtracted from the corresponding weights; and (3) Mean error adjustment case~2, where the mean error of each filter channel is subtracted from the corresponding weights.  \textit{Fig.~\ref{fig:Weights_Noise_Mean_Error_Impact} shows that, among the three cases, intra-channel adjustment (i.e., Mean error adjustment case~2) leads to highest compensation and thereby best results.} 

\section{Type of Non-Uniform Quantization and Design of Supporting DNN Hardware}
\label{sec:novel_number_representation}

Strategy 1 in Section~\ref{sec:analysis_and_strategies} states that aligning quantization scheme with the data distribution helps in restricting the overall quantization error. However, the key challenge is \textit{how can this observation be exploited for improving the efficiency of a DNN-based system.} To address this, we propose \textit{Encoded Low-Precision Binary Signed Digit (ELP\_BSD)} representation that evolves from power-of-two representation and thereby enables the use of shift operations instead of multiplications in processing arrays of DNN accelerators. 
In the following, we discuss the details of our data representation format, starting with the initial concept, limitations of initial proposition, and how ELP\_BSD overcomes these limitations.

\subsubsection{Initial Proposition} 
\begin{figure}[htbp]
\centering
\includegraphics[width=1\linewidth]{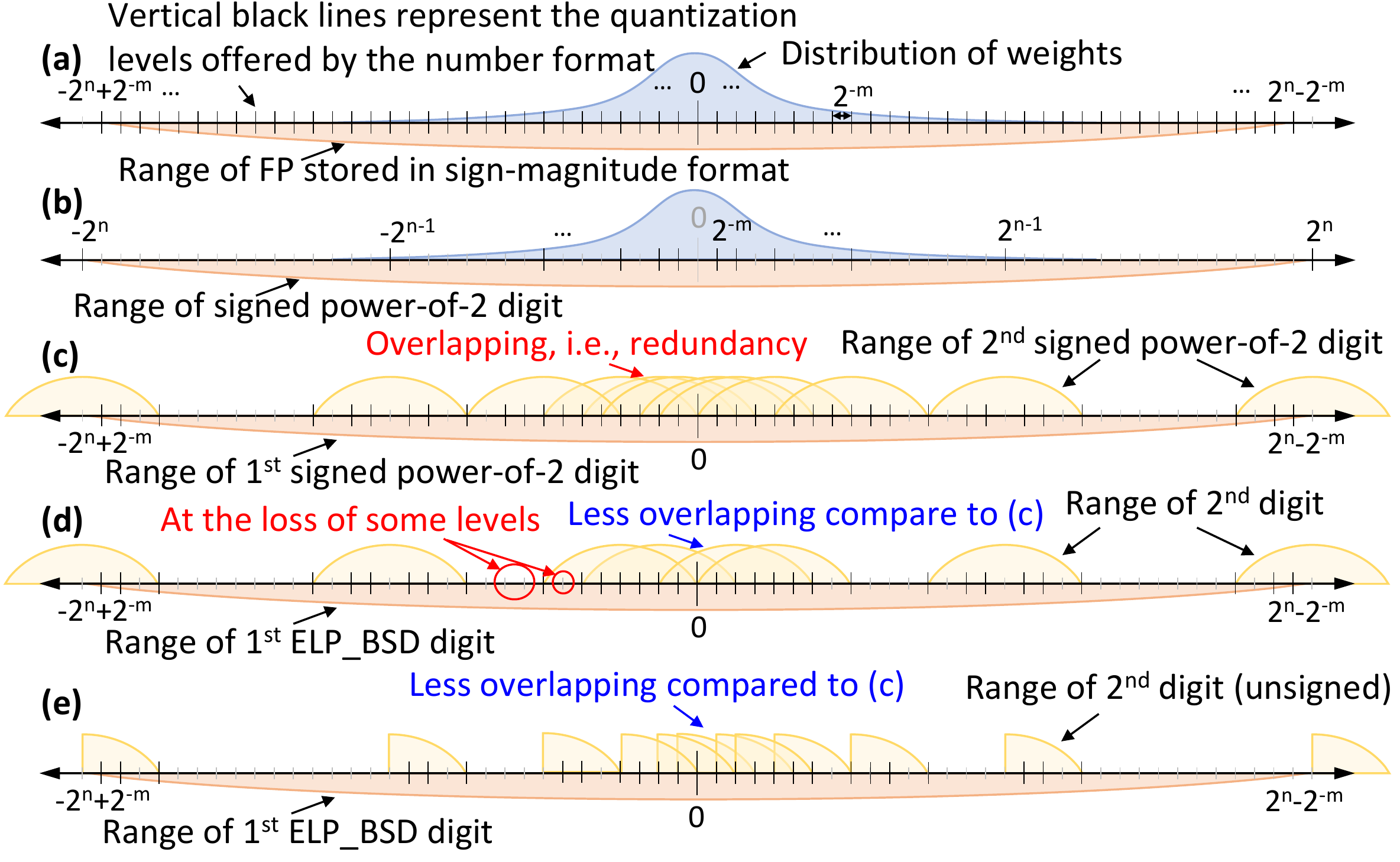}
\caption{Illustration of different data representation formats that show step-by-step evolution of traditional quantization scheme to our ELP\_BSD representation.} 
\label{fig:evolution_of_data_rep_format}
\end{figure}

In this work, the key focus is on exploiting non-uniform quantization for simplifying the MAC unit design as well as in reducing the bit-width of weights, as both contribute towards improving the energy-efficiency. 
Power-of-two quantization is one potential solution, as it allows to replace a costly multiplication operation with a shift operation and reduce the bit-width by storing only the power of 2. 
Moreover, the distribution of quantization levels of power-of-two quantization is aligned with the distribution of DNN weights, as can be observed from Fig.~\ref{fig:evolution_of_data_rep_format}(b). 
However, use of power-of-two quantization results in a significant drop in application-level accuracy due to a significant reduction in the number of quantization levels compared to a traditional FP quantization, which can be observed by comparing Fig.~\ref{fig:evolution_of_data_rep_format}(b) with Fig.~\ref{fig:evolution_of_data_rep_format}(a). \textit{Therefore, almost all the previous works that use power-of-two quantization employ retraining to re-gain the lost accuracy.} 
To offer additional quantization levels to avoid significant accuracy loss due to quantization while benefiting from the advantages of power-of-two quantization scheme, we propose to use sum of signed power-of-two quantization, where more than one signed power-of-two digits are combined to offer additional quantization levels. 
Fig.~\ref{fig:evolution_of_data_rep_format}(c) shows that addition of only a single low-range low-precision signed power-of-two digit can significantly increase the number of unique quantization levels and thereby can help in achieving an ultra-efficient system. 

\subsubsection{Limitations of Initial Proposition and New Improvements} One key issue with sum of signed power-of-two quantization is redundant representations of values. For example, $2^{x_1}-2^{x_2} = 0$ $\forall$ $x_1=x_2$. 
To reduce the amount of redundancy, we propose to reduce the number of possible power of 2 values per digit. 
For example, if a number representation is given as $2^{x_1}-2^{x_2}$ and $x_1, x_2 \in \{0, 1, 2, 3\}$, to reduce the redundancy, we can reduce the set of possible values of $x_1$ to $\{1, 3\}$. 
Fig.~\ref{fig:evolution_of_data_rep_format}(d) shows that reducing the number of possible power of 2 values for the first digit leads to a reduction in the amount of redundancy, which can be observed from the reduced number of yellow semi-circles (representing the range of the second digit) centered at every possible quantization level of the first digit. 
Note that the reduction in redundancy can be exploited to reduce the bit-width of numbers as well as to further simplify the MAC unit design. 
However, a drawback of this is that it can result in a small decrease in the number of quantization levels, as highlighted in Fig.~\ref{fig:evolution_of_data_rep_format}(d).
The redundancy can also be reduced by allowing some digits to have signed values and some only positive (i.e., unsigned) values. This is illustrated in Fig.~\ref{fig:evolution_of_data_rep_format}(e) by restricting the range of the second digit to only positive values. 
\textit{Note that even though the sum of power-of-2 digits leads to some redundancy, it enable us to use shifters instead of multipliers in the hardware, which significantly improves the energy efficiency of DNN systems (as will be highlighted in Section~\ref{sec:results}). 
Moreover, the above explanation shows that the redundancy can be controlled by intelligent selection of the exact number representation format.}

\subsubsection{ELP\_BSD data representation}
\begin{figure}[htbp]
\centering
\includegraphics[width=1\linewidth]{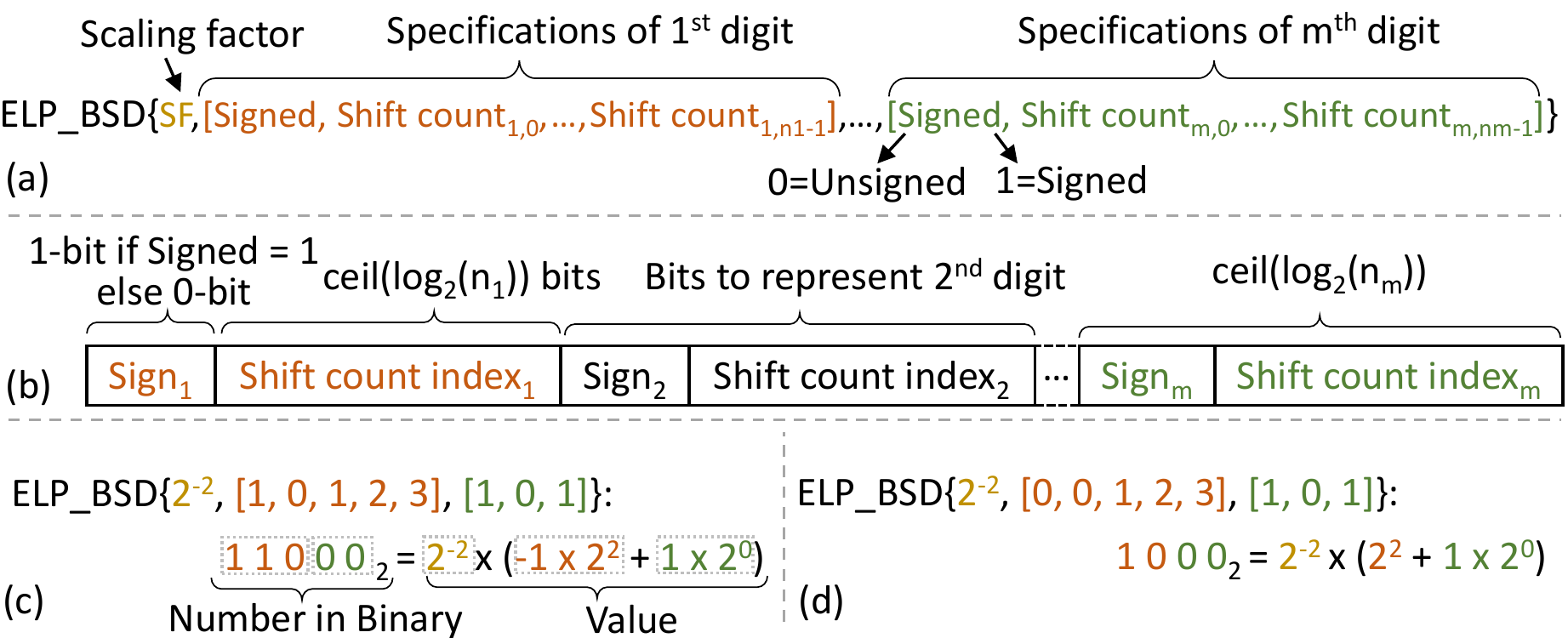}
\caption{(a) Specifications of an ELP\_BSD format. (b) ELP\_BSD format. (c) and (d) Examples to explain conversion between ELP\_BSD format and values.\vspace{-10pt}}
\label{fig:data_rep_format}
\end{figure}

To efficiently represent low-precision sum of signed power-of-two numbers, we define a novel data representation format, Encoded Low-Precision Binary Signed Digit (ELP\_BSD) representation.  Fig.~\ref{fig:data_rep_format}(a) shows how the specifications of a ELP\_BSD representation are defined, and Fig.~\ref{fig:data_rep_format}(b) shows the corresponding binary representation format. As shown in Fig.~\ref{fig:data_rep_format}(b), the bits are divided into $m$ groups, where each group is responsible for representing a single signed power-of-two digit. 
Each group consists of a sign bit and $ceil(log_2(n_i))$ bits to represent the index of shift count in the digit specification, where $n_i$ is the number of different shift counts mentioned for $i^{th}$ digit in the specification. Note that the sign bit is optional and only used when the corresponding digit is signed. Fig.~\ref{fig:data_rep_format} also presents two examples to explain the conversion of ELP\_BSD numbers to FP values. 

\subsubsection{Supporting Hardware Design}
\label{sec:novel_hardware_design}
\begin{figure}[htbp]
\centering
\includegraphics[width=1\linewidth]{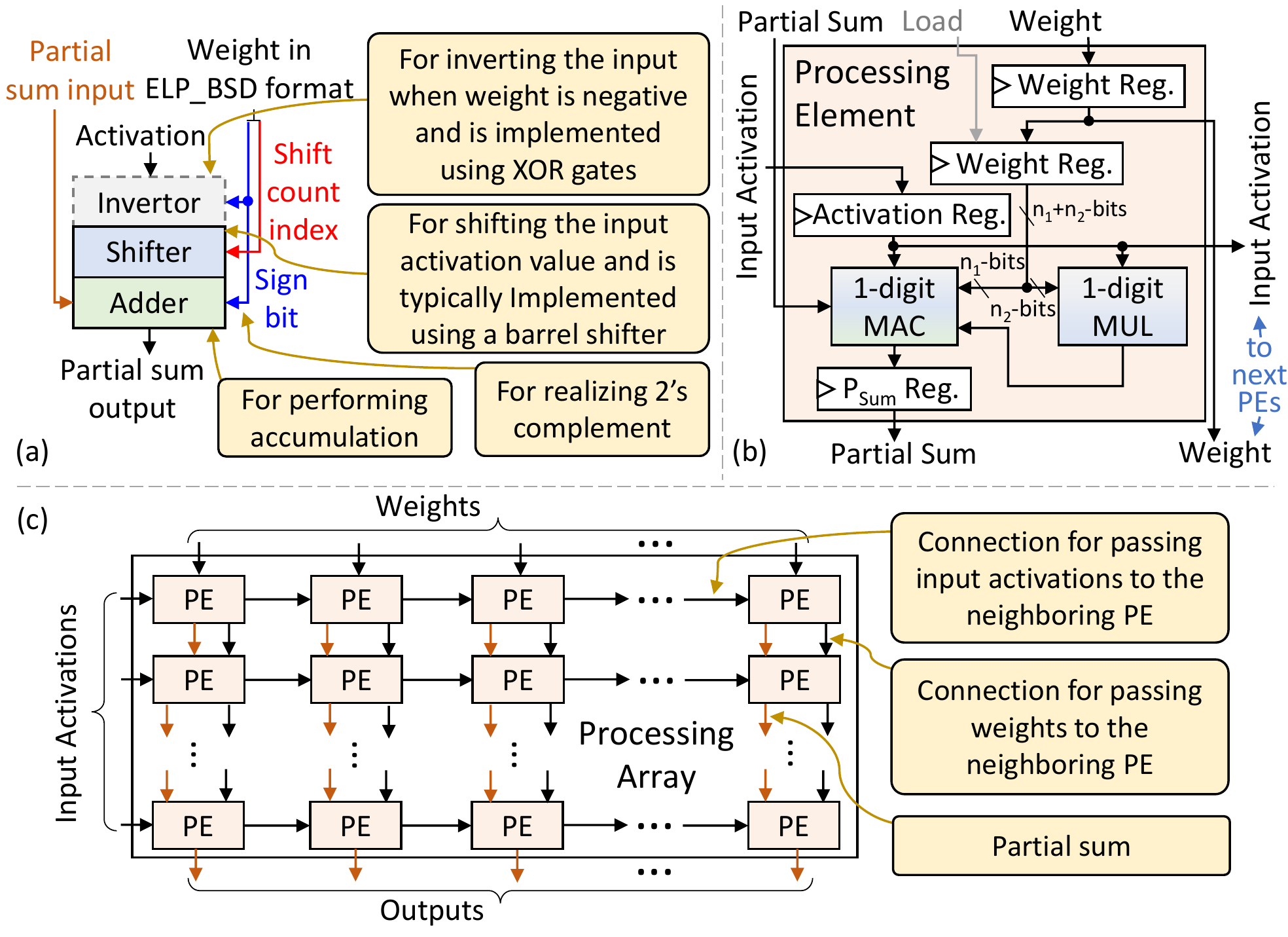}
\caption{(a) Single digit MAC design. (b) Modified processing element for an NPU. (c) A Neural Processing Array architecture.\vspace{-5pt}} 
\label{fig:hardware}
\end{figure}

As shown in Fig.~\ref{fig:data_rep_format}, ELP\_BSD format mainly stores power-of-two digits in an encoded format. To efficiently implement multiplication with a power-of-two digit at hardware-level, shifters (e.g., a barrel shifter) can be used. 
To realize a MAC unit, the shifter is followed by an adder that adds previously computed partial sum to the newly computed product. 
Fig.~\ref{fig:hardware}(a) shows the MAC unit design for the case when weights are represented using a single signed power-of-two digit. 
If we use the same ELP\_BSD format for all the weights, we can hard code the indexing functionality in the shifter and use indexes directly from the encoded weight for multiplication. Moreover, we can choose the set of possible shift counts in a manner that it results in a less complex shifter design. 

In case weights are represented using an ELP\_BSD format that contains multiple digits, we can use multiple of these units (one per-digit) in parallel to compute the partial sums, which then have to be added together to generate one output. 
To achieve this addition, we propose to use a compressor tree followed by a multi-bit adder to add the outputs of the shifters and the partial sum from the previous computation to generate only a single output. 
Fig.~\ref{fig:hardware}(b) shows how multiple single digit MAC units can be integrated in the Processing Elements (PEs) of a Neural Processing Array (NPU), e.g., like the Tensor Processing Unit (TPU)~\cite{Jouppi:2017:IPA:3079856.3080246}. 
Fig.~\ref{fig:hardware}(c) shows the processing array design of the TPU like architecture. This processing array follows a weight stationary dataflow where weights are kept stationary inside the PEs during execution. The input activations are fed from the left and moved towards the right over clock cycles. Similarly, the partial sums are moved towards the bottom of the array. Note that, for this work, we choose to represent activations using FP and 2's complement format, as changing the format of activations won't have any significant impact on the length of the adders in PEs. 
It is also important to highlight here that in most of the case 1-3 single digit MAC units per PE are sufficient to meet the user-defined accuracy constraints.

\section{Our Methodology for Efficient Approximation of CNNs through Non-Uniform Quantization}
\label{sec:novel_methodology_overall}

\begin{figure}[htbp]
\centering
\includegraphics[width=1\linewidth]{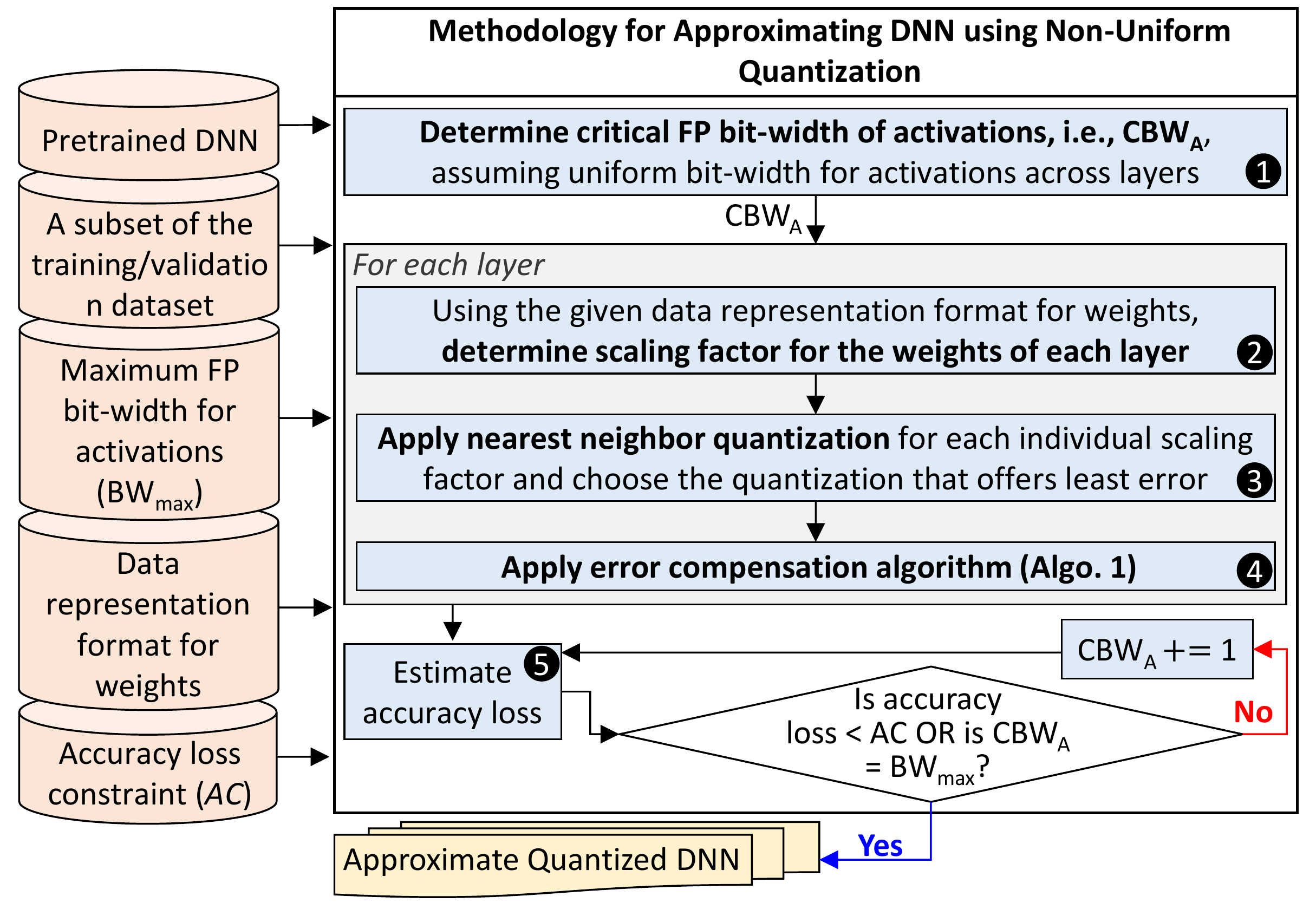}
\caption{Our DNN quantization methodology\vspace{-10pt}}
\label{fig:meth}
\end{figure}

Fig.~\ref{fig:meth} presents our methodology for approximating CNNs through non-uniform quantization while exploiting the strategies mentioned in Section~\ref{sec:analysis_and_strategies}. The following steps explain working of the methodology.

\begin{enumerate}[leftmargin=*]
    \item \textbf{Determine critical FP bit-width of activations:} Starting from maximum allowed FP bit-width activations, we gradually reduce the bit-width to find the critical point after which the accuracy loss of the input DNN raises above the user-defined accuracy loss constraint (AC). For this step, we assume bit-width of all the activations to be the same. The key intuition behind this step is that a decrease in the activations' bit-width results in a linear decrease in the width of MAC units, which can help in improving the energy-efficiency. This step outputs critical bit-width for activations, denoted as $CBW_A$.
    
    \item \textbf{Determine scaling factor for weights:} Given the data representation format for weights, we compute the scaling factor for weights of each layer of the input DNN separately. For this work, in case of ELP\_BSD format, we compute the scaling factor as $SF = max(weights)/2^{max(shift\_ counts)}$.
    
    \item \textbf{Apply nearest neighbor quantization:} Using the data representation format and the scaling factors, we generate a table of possible quantization levels (TQL) for each layer of the input DNN. Then to perform quantization, we replace the weights with their corresponding nearest values in the tables. 
    
    \item \textbf{Apply error compensation algorithm:} For each convolutional layer, we pass the weights to Algo.~\ref{Algorithm:Error_Compensation}, which (partially) compensates for errors introduced due to quatization of weights by exploiting Strategy 2 mentioned in Section~\ref{sec:analysis_and_strategies}. Note that as most of state-of-the-art architectures use small filter sizes, e.g., 3x3, Algo.~\ref{Algorithm:Error_Compensation} focuses on compensating the overall channel-level mean quantization error in filter weights. To achieve this, it computes the mean quantization error of a channel of a filter, locates the values that can be mapped to their other neighboring quantization level to reduce the mean error, sorts all the located values based on a cost function and starts altering the values starting form the values having least cost till the point the absolute mean error is decreasing. 
    
    \item \textbf{Estimate the overall accuracy loss:} In this step, we compute the accuracy to check if the user-defined accuracy constraint is met. In case the constraint is not satisfied, the algorithm increases the value of $CBW_A$ by 1 and performs accuracy evaluation again. If the constraint is still not met and $CBW_A$ becomes equal to $BW_{max}$, it outputs the latest quantized DNN. 
\end{enumerate}

\setlength{\textfloatsep}{0pt}
\begin{algorithm}[ht]
\SetAlgoLined
\small
\textbf{Input:} Un-quantized weights of a CONV layer ($UnQunat\_W$); Table of possible quantization levels ($TQL$)

\KwResult{Quantized weights ($Quant\_W$)}

$Quant\_W$ $\leftarrow$ NNQuat($UnQunat\_W$, $TQL$); \% Nearest neighbor quantization \\
$Error\_W$ $\leftarrow$ $UnQunat\_W$ – $Quant\_W$; \% Error in weights

\For{$(i = 1;\ i \leq$ No of filters $;\ i += 1)$}{
    \For{$(j = 1;\ i \leq$ No of channels $;\ j += 1)$}{
        $Mean\_Err$ = mean($Error\_W$( :, :, $i$, $j$)); \% Mean error of the channel\\
        $S$ $\leftarrow$ Subset of $UnQunat\_W$( :, :, $i$, $j$) having corresponding error opposite in sign to $Mean\_Err$\\
        $SO$ $\leftarrow$ Values of $S$ quantized to closet levels in the opposite direction to the nearest neighbor\\
        $Cost_S$ = $abs(S-SO)$\\
        $Sorted\_S$ $\leftarrow$ Set of sorted values of $S$ in order of increasing cost\\
        \For{$(k = 1;\ i \leq$ No of values in $Sorted\_S$ $;\ k += 1)$}{
            $New_Mean_Err$ $\leftarrow$ Mean quantization error if the quantized value of $Sorted\_S(k)$ in $Quant\_W$ is replaced with the corresponding value from $SO$\\
            \eIf{abs(Mean\_Err) > abs(New\_Mean\_Err)}{
                Accept the change in quantization level of value corresponding to $Sorted\_S(k)$ in $Quant\_W$\\
                $Mean\_Err  = New\_Mean\_Err$
            }{
                break
            }
        }
    }
}
 \caption{Pseudo-code for error compensation}
 \label{Algorithm:Error_Compensation}
\end{algorithm}

\section{Results and Discussion}
\label{sec:results}
\subsection{Experimental Setup}
To evaluate CoNLoCNN, we extended MatConvNet~\cite{vedaldi15matconvnet} framework for FP implementation and our CoNLoCNN methodology. We evaluated CoNLoCNN using two popular DNNs used for benchmarking FP implementations, i.e., AlexNet and VGG-16 trained on the ImageNet dataset. 
For hardware synthesis, we implemented PEs composed of different MAC unit designs in Verilog and synthesized for the TSMC 65nm technology using Cadence Genus. 

\subsection{Effectiveness of Our Error Compensation Strategy, i.e., Algorithm~\ref{Algorithm:Error_Compensation}}
\begin{figure}[htbp]
\centering
\includegraphics[width=1\linewidth]{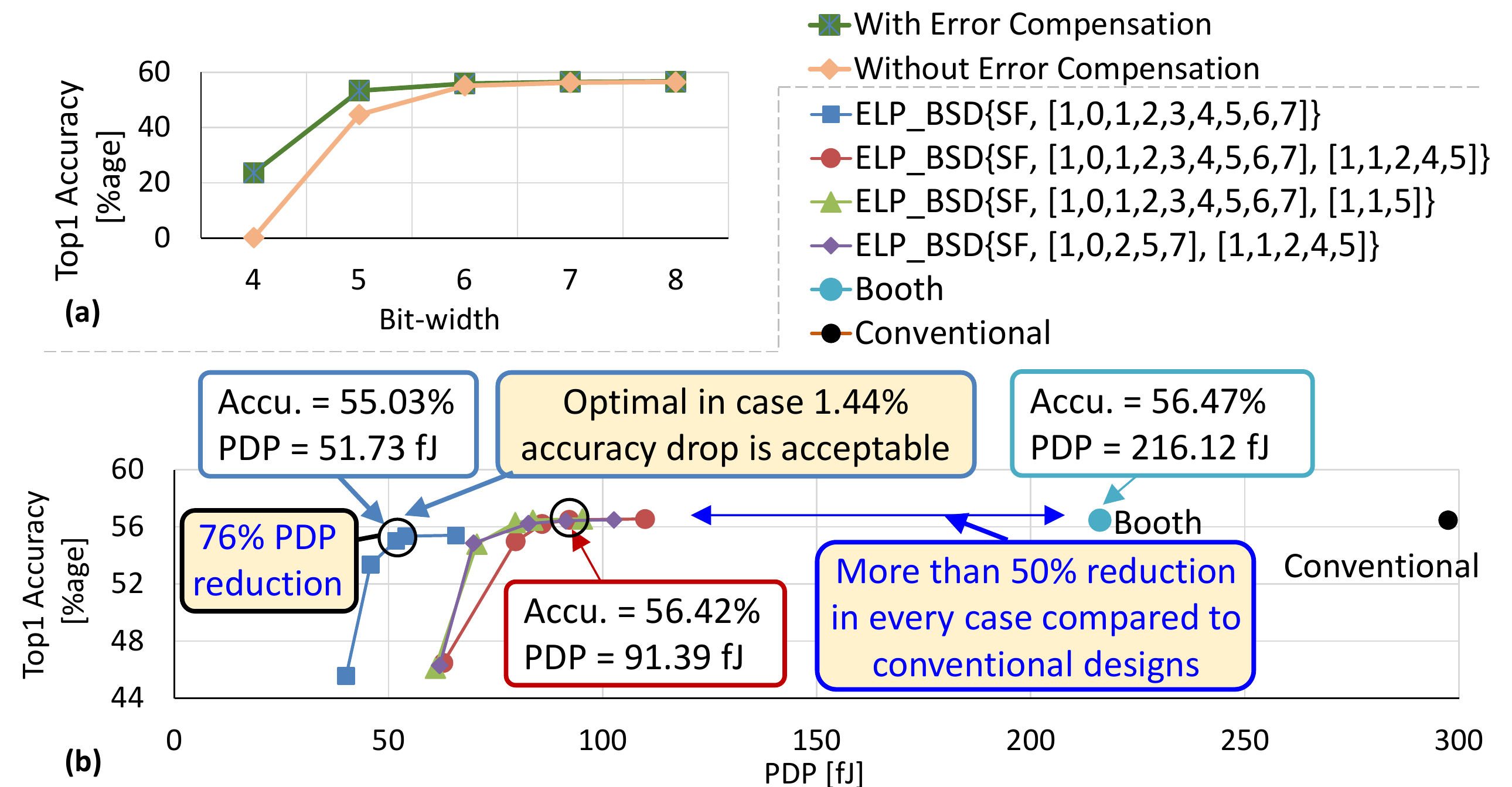}
\caption{(a) Effectiveness of our error compensation strategy when used with traditional FP quatization for the AlexNet. (b) Accuracy of the AlexNet vs. PDP for different ELP\_BSD data representations.\vspace{10pt}}
\label{fig:Soft_results}
\end{figure}

To demonstrate the effectiveness of error compensation algorithm, we applied it with FP implementation and compared the results with conventional FP implementation. Note that for this experiment, we assumed the bit-width of weights and activations to be the same and uniform across all the layers of the DNN. Fig.~\ref{fig:Soft_results}(a) shows the results for the AlexNet. As shown in the figure, our error compensation strategy helps in improving the accuracy of the network, specifically at lower bit-widths. 

\subsection{Effectiveness of CoNLoCNN for State-of-the-Art DNNs}
To demonstrate the effectiveness of our overall methodology, we considered four different ELP\_BSD data representation formats. The formats are listed in Table~\ref{tab:hardware_table} along with the hardware characteristics of the PEs implemented using the corresponding MAC designs for a 32x32 processing array (shown in Fig.~\ref{fig:hardware}(c)). Note that the scaling factor (represented by $x$ in the table) of the representations is not considered to be the same across layers, and it is selected based on the statistics of the parameters of the corresponding layer. 
For each representation, we considered five different activation bit-widths, i.e., 8-bit till 4-bit, to study the impact of change in activation bit-width on the accuracy of the network and the hardware efficiency. 
Fig.~\ref{fig:Soft_results}(b) shows the accuracy vs. PDP results achieved when CoNLoCNN is used for the AlexNet considering the ELP\_BSD configurations mentioned in Table~\ref{tab:hardware_table}. The plots on the left inside Fig.~\ref{fig:Soft_results}(b) are generated by CoNLoCNN while the two points on the right represent conventional designs considered for comparison. The plot shows that as the bit-width of the activations decreases, we observe a slight decrease till a point after which the rate of decay increases drastically. 
Moreover, different ELP\_BSD formats offer different accuracy-efficiency characteristics. 
\textit{The key thing to observe in Fig.~\ref{fig:Soft_results}(b) is that even the most power consuming PE design generated by the proposed CoNLoCNN framework offers around 50\% reduction in PDP compared to conventional designs.} In case 1.44\% drop in accuracy is acceptable, the proposed method can offer around 76\% reduction in the PDP. Similar results are observed for the VGG-16 network. 

\begin{table}[h] 
\centering
\caption{Hardware characteristics of the PEs designed using our methodology for some of ELP\_BSD representations and their comparison with booth multiplier-based and conventional multiplier-based PEs.} 
\includegraphics[width=1\linewidth]{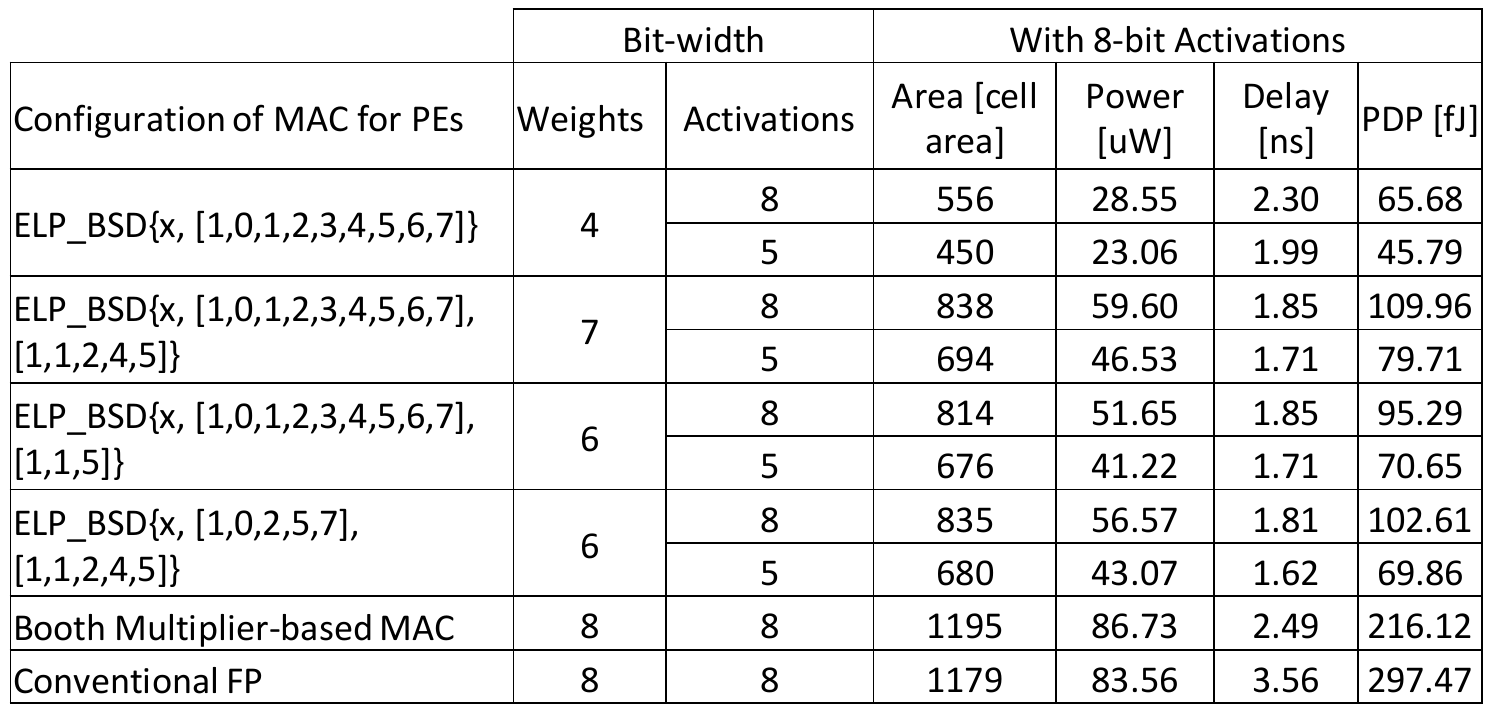}
\vspace{-20pt}
\label{tab:hardware_table}
\end{table}

\subsection{Comparison with the state-of-the-art}
To compare the results with the state-of-the-art, we implemented CAxCNN~\cite{9137167} inside our framework. We selected Canonical Approximate (CA) representation with 1 non-zero digit, where the weights are converted using their exhaustive search algorithm (i.e., their best algorithm) to have a fair comparison. For AlexNet, CAxCNN with 1 non-zero digit CA representation achieves 50.9\% top1 accuracy while CoNLoCNN achieves 55.4\% accuracy (i.e., close to the baseline). This improvement is mainly due to our error compensation strategy. The quantization levels offered by 1 non-zero digit CA are almost the same as offered by ELP\_BSD\{SF,[1,0,1,2,3,4,5,6,7]\} format with the only difference being of '0', which is not present in ELP\_BSD\{SF,[1,0,1,2,3,4,5,6,7]\}. However, absence of '0' does not affect the accuracy due to the presence of 1 and -1 quantization levels and the use error compensation. Note that the absence of '0' in the highlighted case helps in achieving a simplified PE design. 
Even in the best possible scenario, CA representation would require 5~bits per weight while ELP\_BSD\{SF,[1,0,1,2,3,4,5,6,7]\} requires 4~bits per weight. Similar to the AlexNet, for the VGG-16, we observed 3\% higher accuracy with CoNLoCNN compared to CAxCNN. Note that CoNLoCNN not only helps in reducing the complexity of the hardware but also helps in reducing the memory footprint of DNNs unlike other approximate computing works (e.g.,~\cite{mrazek2019alwann}) that operate at 8-bit precision. 

\section{Conclusion}

In this paper, we proposed CoNLoCNN, a framework to enable energy-efficient low-precision approximate DNN inference. CoNLoCNN mainly exploits non-uniform quantization of weights to simplify processing elements in the computational array of DNN accelerators and correlation between activation values to (partially) compensate for the quantization errors without any run-time overheads. We also proposed, Encoded Low-Precision Binary Signed Digit (ELP\_BSD) representation, to reduce the bit-width of weights while ensuring direct use of the encoded weight in computations by designing supporting MAC units.

\section*{Acknowledgment}

This research is partly supported by the ASPIRE AARE Grant (S1561) on "Towards Extreme Energy Efficiency through Cross-Layer Approximate Computing".

\bibliographystyle{IEEEtran}
\bibliography{Refs}

\end{document}